\documentclass[
    ,final            
  ]
  {aipproc}

\layoutstyle{6x9}

\begin{document}

\title{Scalar mesons in a linear sigma model with (axial-)vector mesons}


\author{D. Parganlija}{
  address={Institute for Theoretical Physics, Vienna University of Technology,
Wiedner Hauptstr.\ 8-10, A--1040 Vienna, Austria}
}
\author{P. Kov\'acs}{
  address={Institute for Particle and Nuclear Physics, Wigner Research Center 
for Physics, Hungarian Academy of Sciences, H-1525 Budapest, Hungary}
}
\author{Gy. Wolf}{
  address={Institute for Particle and Nuclear Physics, Wigner Research Center 
for Physics, Hungarian Academy of Sciences, H-1525 Budapest, Hungary}
}

\author{F. Giacosa}{
  address={Institute for Theoretical Physics, Johann Wolfgang Goethe University,
Max-von-Laue-Str.\ 1, D-60438 Frankfurt am Main, Germany}
}
\author{D.H. Rischke}{
  address={Institute for Theoretical Physics, Johann Wolfgang Goethe University,
Max-von-Laue-Str.\ 1, D-60438 Frankfurt am Main, Germany}
}

\begin{abstract}
The structure of the scalar mesons has been a subject of debate for many 
decades. In this work we look for $\bar{q}q$ states among the physical 
resonances using an extended Linear Sigma Model that contains scalar, 
pseudoscalar, vector, and axial-vector mesons both in the non-strange and 
strange sectors. We perform global fits of meson masses, decay widths and 
amplitudes in order to ascertain whether the scalar $\bar{q}q$ states are 
below or above 1 GeV. We find the scalar states above 1 GeV to be preferred 
as $\bar{q}q$ states.
\end{abstract}

\maketitle


\section{Introduction}

Understanding the meson mass spectrum in the region below 2 GeV
is one of the fundamental problems of QCD. While the quark model 
seems to work very well for many resonances, some fundamental questions, 
such as the constituent quark content of scalar and axial-vector 
resonances, are still unanswered.
The Particle Data Group (PDG) \cite{PDG} suggests in the region below 
1.8 GeV the existence of five $I(J^{PC})=0(0^{++})$ states: 
$f_{0}(500)$, $f_{0}(980)$, $f_{0}(1370)$, $f_{0}(1500)$, and $f_{0}(1710)$; 
two $I=1$ states: $a_{0}(980)$ and $a_{0}(1450)$; and two $I=1/2$ states:
$K_{0}^{\star}(800)$ (or $\kappa$) and $K_{0}^{\star}(1430)$.
A description of all mentioned scalar states as $\bar{q}q$ states is not 
possible, since the number of physical resonances is much larger than 
the number of resonances that can be constructed within a $\bar{q}q$ 
picture of mesons, e.g. in the isoscalar sector two states compared to five 
experimentally seen resonances. 
The question is: which of the five states are (predominantly)
quarkonia?

Understanding these issues is not only crucial for hadron vacuum spectroscopy
but is also important at nonzero temperatures and densities, because the
correct identification of the chiral partner of the pion and of the $\rho$
meson is necessary for a proper description of the in-medium properties of 
hadrons \cite{Achim}.
The answers to this fundamental question is in principle contained in
the QCD Lagrangian. Unfortunately, QCD cannot be solved by analytic
means from first principles in the low-energy domain. 
For this reason, effective theories have been developed which share some of
the underlying symmetries of QCD. The QCD Lagrangian
exhibits, in addition to the local $SU(3)_{c}$ color symmetry and the discrete
$CPT$ symmetry, a global $U(N_{f})_{L}\times U(N_{f})_{R}\equiv
U(1)_{V}\times U(1)_{A}\times SU(N_{f})_{V}\times SU(N_{f})_{A}$ chiral
symmetry which is broken in several ways: spontaneously [due to the chiral
condensate $\langle\bar{q}q\rangle=\langle\bar{q}_{R}q_{L}+\bar{q}_{L}%
q_{R}\rangle\neq0$], explicitly (due to non-vanishing quark
masses), as well as at the quantum level [the $U(1)_{A}$ anomaly].

In the framework of effective theories the chiral symmetry of QCD can be
realized along two lines: linearly \cite{gellmanlevy} and non-linearly
\cite{weinberg}. In this contribution, we present a linear sigma model 
containing scalar, pseudoscalar, vector, and axial-vector mesons with both 
non-strange and strange quantum numbers. In view of the large number of the 
fields involved, our model shall be referred to as the 
``extended Linear Sigma Model'', or ``eLSM''.

\section{THE MODEL}

The Lagrangian of our model \cite{eLSM} reads
\begin{eqnarray}
\mathcal{L}  &
=&\mathop{\mathrm{Tr}}[(D_{\mu}\Phi)^{\dagger}(D_{\mu}\Phi)]-m_{0}
^{2}\mathop{\mathrm{Tr}}(\Phi^{\dagger}\Phi)-\lambda_{1}
[\mathop{\mathrm{Tr}}(\Phi^{\dagger}\Phi)]^{2}-\lambda_{2}
\mathop{\mathrm{Tr}}(\Phi^{\dagger}\Phi)^{2}{\nonumber}\\
&&  -\frac{1}{4}\mathop{\mathrm{Tr}}(L_{\mu\nu}^{2}+R_{\mu\nu}^{2}
)+\mathop{\mathrm{Tr}}\left[  \left(  \frac{m_{1}^{2}}{2}+\Delta\right)
(L_{\mu}^{2}+R_{\mu}^{2})\right]
+\mathop{\mathrm{Tr}}[H(\Phi+\Phi^{\dagger})]{\nonumber}\\
& & +c_{1}(\det\Phi-\det\Phi^{\dagger})^{2}+i\frac{g_{2}}{2}
(\mathop{\mathrm{Tr}}\{L_{\mu\nu}[L^{\mu},L^{\nu}
]\}+\mathop{\mathrm{Tr}}\{R_{\mu\nu}[R^{\mu},R^{\nu}]\}){\nonumber}\\
&&  +\frac{h_{1}}{2}\mathop{\mathrm{Tr}}(\Phi^{\dagger}\Phi
)\mathop{\mathrm{Tr}}(L_{\mu}^{2}+R_{\mu}^{2})+h_{2}
\mathop{\mathrm{Tr}}[(L_{\mu}\Phi)^{2}+(\Phi R_{\mu} )^{2}]+2h_{3}
\mathop{\mathrm{Tr}}(L_{\mu}\Phi R^{\mu}\Phi^{\dagger}).{\nonumber}\\
&&  +g_{3}[\mathop{\mathrm{Tr}}(L_{\mu}L_{\nu}L^{\mu}L^{\nu}
)+\mathop{\mathrm{Tr}}(R_{\mu}R_{\nu}R^{\mu}R^{\nu})]+g_{4}
[\mathop{\mathrm{Tr}}\left(  L_{\mu}L^{\mu}L_{\nu}L^{\nu}\right)
+\mathop{\mathrm{Tr}}\left(  R_{\mu}R^{\mu}R_{\nu}R^{\nu}\right)
]{\nonumber}\\
&&  +g_{5}\mathop{\mathrm{Tr}}\left(  L_{\mu}L^{\mu}\right)
\,\mathop{\mathrm{Tr}}\left(  R_{\nu}R^{\nu}\right)  +g_{6}
[\mathop{\mathrm{Tr}}(L_{\mu}L^{\mu})\,\mathop{\mathrm{Tr}}(L_{\nu}L^{\nu
})+\mathop{\mathrm{Tr}}(R_{\mu}R^{\mu})\,\mathop{\mathrm{Tr}}(R_{\nu}R^{\nu})]
 , \label{eq:Lagrangian}%
\end{eqnarray}
where
\begin{eqnarray}
D^{\mu}\Phi &  \equiv &\partial^{\mu}\Phi-ig_{1}(L^{\mu}\Phi-\Phi R^{\mu
})-ieA^{\mu}[T_{3},\Phi]\;,\nonumber\\
L^{\mu\nu}  &  \equiv&\partial^{\mu}L^{\nu}-ieA^{\mu}[T_{3},L^{\nu}]-\left\{
\partial^{\nu}L^{\mu}-ieA^{\nu}[T_{3},L^{\mu}]\right\}\;  ,\nonumber\\
R^{\mu\nu}  &  \equiv&\partial^{\mu}R^{\nu}-ieA^{\mu}[T_{3},R^{\nu}]-\left\{
\partial^{\nu}R^{\mu}-ieA^{\nu}[T_{3},R^{\mu}]\right\}
\; ,\nonumber
\end{eqnarray}
The quantities $\Phi$, $R^{\mu}$, and $L^{\mu}$ represent
the scalar and vector nonet:
\begin{eqnarray}
\Phi &  = &\sum_{i=0}^{8}(S_{i}+iP_{i})T_{i}=\frac{1}{\sqrt{2}}\left(
\begin{array}
[c]{ccc}%
\frac{(\sigma_{N}+a_{0}^{0})+i(\eta_{N}+\pi^{0})}{\sqrt{2}} & a_{0}^{+}
+i\pi^{+} & K_{0}^{\star+}+iK^{+}\\
a_{0}^{-}+i\pi^{-} & \frac{(\sigma_{N}-a_{0}^{0})+i(\eta_{N}-\pi^{0})}
{\sqrt{2}} & K_{0}^{\star0}+iK^{0}\\
K_{0}^{\star-}+iK^{-} & {\bar{K}_{0}^{\star0}}+i{\bar{K}^{0}} & \sigma
_{S}+i\eta_{S}%
\end{array}
\right)  ,\nonumber\\
L^{\mu}  &  =&\sum_{i=0}^{8}(V_{i}^{\mu}+A_{i}^{\mu})T_{i}=\frac{1}{\sqrt{2}
}\left(
\begin{array}
[c]{ccc}%
\frac{\omega_{N}+\rho^{0}}{\sqrt{2}}+\frac{f_{1N}+a_{1}^{0}}{\sqrt{2}} &
\rho^{+}+a_{1}^{+} & K^{\star+}+K_{1}^{+}\\
\rho^{-}+a_{1}^{-} & \frac{\omega_{N}-\rho^{0}}{\sqrt{2}}+\frac{f_{1N}
-a_{1}^{0}}{\sqrt{2}} & K^{\star0}+K_{1}^{0}\\
K^{\star-}+K_{1}^{-} & {\bar{K}}^{\star0}+{\bar{K}}_{1}^{0} & \omega
_{S}+f_{1S}%
\end{array}
\right)  ^{\mu},\nonumber\\
R^{\mu}  &  =&\sum_{i=0}^{8}(V_{i}^{\mu}-A_{i}^{\mu})T_{i}=\frac{1}{\sqrt{2}
}\left(
\begin{array}
[c]{ccc}%
\frac{\omega_{N}+\rho^{0}}{\sqrt{2}}-\frac{f_{1N}+a_{1}^{0}}{\sqrt{2}} &
\rho^{+}-a_{1}^{+} & K^{\star+}-K_{1}^{+}\\
\rho^{-}-a_{1}^{-} & \frac{\omega_{N}-\rho^{0}}{\sqrt{2}}-\frac{f_{1N}%
-a_{1}^{0}}{\sqrt{2}} & K^{\star0}-K_{1}^{0}\\
K^{\star-}-K_{1}^{-} & {\bar{K}}^{\star0}-{\bar{K}}_{1}^{0} & \omega
_{S}-f_{1S}%
\end{array}
\right)  ^{\mu}, \nonumber%
\end{eqnarray}
where the assignment to physical particles is shown as well\footnote{With the 
exception of the $(0-8)$ sector where
particle mixing takes place.}.  Here,
$T_{i}\,(i=0,\ldots,8)$ denote the generators of $U(3)$, while $S_{i}$ represents
the scalar, $P_{i}$ the pseudoscalar, $V_{i}^{\mu}$ the vector, and
$A_{i}^{\mu}$ the axial-vector meson fields, and $A^{\mu}$ is the
electromagnetic field. It should be noted that here and below 
we use the so-called non strange -- strange basis in the
$(0-8)$ sector, defined as
\begin{equation}
\varphi_{N}   = \frac{1}{\sqrt{3}} \left( \sqrt{2}\; \varphi_{0}+
\varphi_{8}\right)\;,\qquad
\varphi_{S}  =  \frac{1}{\sqrt{3}} \left(
  \varphi_{0}-\sqrt{2}\;\varphi_{8} \right) \;,\quad\quad
\varphi\in(S_i,P_i,V_i^{\mu},A_i^{\mu})\;, \label{eq:nsbase}%
\end{equation}
which is more suitable for our calculations. Moreover, 
$H$ and $\Delta$ are constant external fields defined as
\begin{eqnarray}
H  &  = &H_{0}T_{0}+H_{8}T_{8}=\left(
\begin{array}
[c]{ccc}%
\frac{h_{0N}}{2} & 0 & 0\\
0 & \frac{h_{0N}}{2} & 0\\
0 & 0 & \frac{h_{0S}}{\sqrt{2}}%
\end{array}
\right)\;  ,\label{eq:expl_sym_br_epsilon}\\
\Delta &  =&\Delta_{0}T_{0}+\Delta_{8}T_{8}=\left(
\begin{array}
[c]{ccc}%
\frac{\tilde{\delta}_{N}}{2} & 0 & 0\\
0 & \frac{\tilde{\delta}_{N}}{2} & 0\\
0 & 0 & \frac{\tilde{\delta}_{S}}{\sqrt{2}}%
\end{array}
\right)  \equiv\left(
\begin{array}
[c]{ccc}%
\delta_{N} & 0 & 0\\
0 & \delta_{N} & 0\\
0 & 0 & \delta_{S}%
\end{array}
\right)  \quad . \label{eq:expl_sym_br_delta}%
\end{eqnarray}
Throughout this
work we assume exact isospin symmetry
for $u$ and $d$ quarks, such that the
first two diagonal elements in Eqs.\ \eqref{eq:expl_sym_br_epsilon} and
\eqref{eq:expl_sym_br_delta} are identical.

All fields in our model represent $\bar{q} q$ states, as discussed in 
Ref.\ \cite{Paper1}.
In the non-strange sector, we assign the fields $\vec{\pi}$ and $\eta_{N}$ to
the pion and the non-strange part of the $\eta$ and $\eta'$ mesons, 
$\eta_{N}\equiv
(\bar{u}u+\bar{d}d)/\sqrt{2}$. The fields $\omega_{N}^{\mu}$ and $\vec{\rho
}^{\;\mu}$ represent the $\omega(782)$ and $\rho(770)$ vector mesons,
respectively, and the fields $f_{1N}^{\mu}$ and $\vec{a}_{1}^{\;\mu}$ represent
the $f_{1}(1285)$ and $a_{1}(1260)$ mesons, respectively. In the strange
sector, we assign the $K$ fields to the kaons; the $\eta_{S}$ field is the
strange contribution to the physical $\eta$ and $\eta^{\prime}$ fields
[$\eta_{S}\equiv\bar{s}s$]; the $\omega_{S}$, $f_{1S}$, $K^{\star}$, and
$K_{1}$ fields correspond to the $\phi(1020)$, $f_{1}(1420)$, $K^{\star}%
(892)$, and $K_{1}(1270)$ [or $K_{1}(1400)$] mesons, respectively.

Unfortunately, the assignment of the scalar fields is substantially less
clear. Experimental data suggest existence of five scalar-isoscalar states
below $1.8$ GeV: $f_{0}(500)$, $f_{0}(980)$, $f_{0}(1370)$,
$f_{0}(1500)$, and $f_{0}(1710)$, 
all of which could, in principle, be candidates for $f_{0}^{L}$
and $f_0^H$.
Similarly, the isospin triplet $\vec{a}_{0}$ can be assigned to different
physical resonances -- although, in this case, there are only two candidate
states: $a_{0}(980)$ and $a_{0}(1450)$. An analogous statement holds for the
scalar kaon $K_{0}^{\star}$ that can be assigned to the resonances
$K_{0}^{\star}(800)$ or $K_{0}^{\star}(1430)$.

In the Lagrangian (\ref{eq:Lagrangian}) there are two terms which break the
original $U(3)_{L}\times U(3)_{R}$ [$=U(3)_{A}\times U(3)_{V}$]
symmetry, 
namely the determinant term, which
breaks the $U(1)_{A}$ symmetry, and the explicit symmetry breaking terms
of Eqs.~\eqref{eq:expl_sym_br_epsilon} and (\ref{eq:expl_sym_br_delta}), which
break $U(3)_{A}$, if $H_{0}, \Delta_0 \neq0$ and $U(3)_{V}\rightarrow
SU(2)_{V}\times U(1)_{V}$, if in addition
$H_{8}, \Delta_8 \neq 0$ [for more details
see, e.g.\ Ref.\ \cite{Lenaghan:2000ey}]. Besides this explicit symmetry
breaking the chiral symmetry is also broken
spontaneously. If isospin symmetry is exact, only the $\sigma
_{N}$ and $\sigma_{S}$ scalar fields, carrying the same quantum
numbers as the vacuum, can have nonzero vacuum expectation
values (vev's)\footnote{In case of isospin breaking, also
$\sigma_{3}$ would have a nonzero vev.}. Moreover, the parameter $c_{1}$
describes the axial anomaly, i.e., the $U(1)_{A}$ symmetry is explicitly
broken by this term.

After spontaneous symmetry breaking, the fields with nonzero vev's are 
shifted by their expectation values, namely, 
$\sigma_{N/S}\rightarrow\sigma_{N/S}+\phi_{N/S}$, where we have introduced 
$\phi_{N/S}\equiv \langle \sigma_{N/S} \rangle$. 
After substituting the shifted fields into the Lagrangian
\eqref{eq:Lagrangian}, one obtains the tree-level
masses by selecting all terms quadratic in the fields.
The (square) mass matrices are in general non-diagonal due to the mixing 
among particles sitting in the center of a given nonet. Besides the mixing 
inside the nonets there are other terms which mix different nonets 
because of the vacuum expectation values of the $\sigma$ fields. The mass 
matrices can be diagonalized by appropriate orthogonal transformations. 
In order to retain the canonical normalization for the fields, one has to 
introduce wavefunction renormalization constants, too.

From the Lagrangian one can also derive several decay widths. The mass and 
decay-width formulas can be found in Ref.~\cite{eLSM}.

\section{Results}

\label{ssec:SSB_and_mass}

\begin{figure}
  \includegraphics[height=.28\textheight]{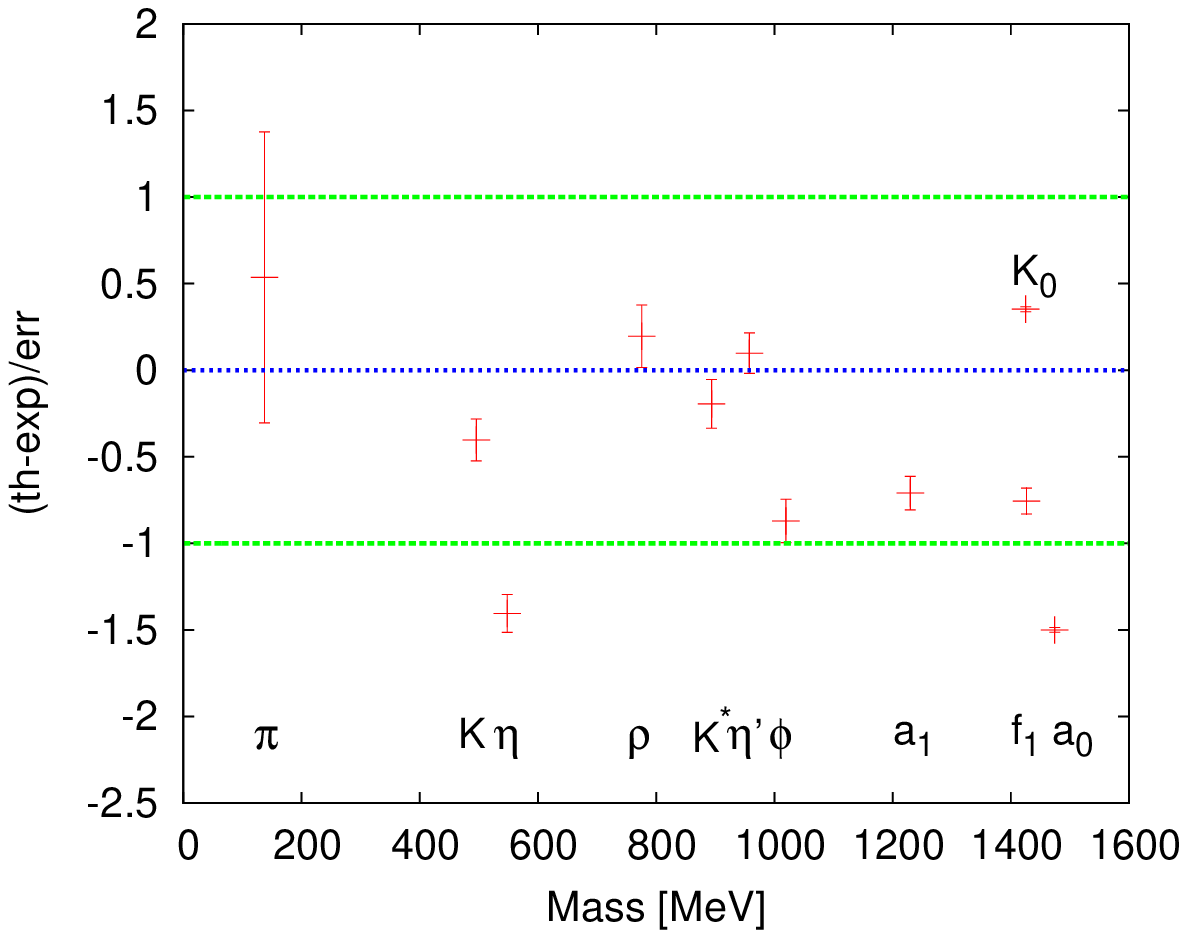}
  \includegraphics[height=.28\textheight]{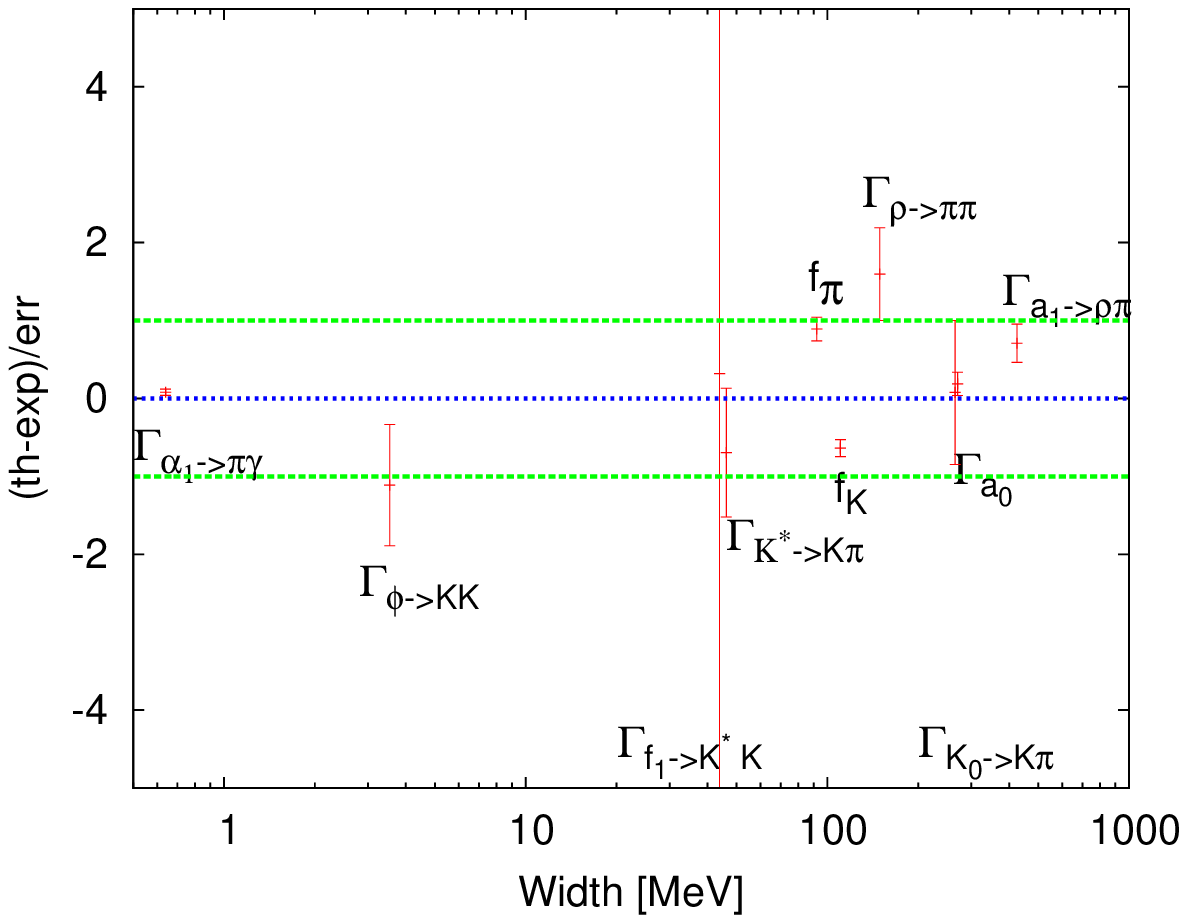}
  \caption{Fits to masses and decay widths. 
Shown is the difference of the theoretical and 
experimental values divided by the experimental errors.
The errorbars correspond to the theoretical errors as emerging from
our fit. }
 \label{Fit}
\end{figure}

We perform a global fit of the parameters of the Lagrangian
\eqref{eq:Lagrangian} 
to experimentally known quantities 
(decay constants, masses, and decay widths as well as
amplitudes), for details see Ref.~\cite{eLSM}. 
We do not include the scalar-isoscalar states into 
the fit, but we test \textit{a posteriori} their assignment and
phenomenology. We do, however, include the isotriplet and isodoublet 
quark-antiquark scalar states and we
test all four combinations $a_{0}(1450)~/~K_{0}^{\star}(1430)$,
$a_{0}(980)~/~K_{0}^{\star}(800)$, $a_{0}(980)~/~K_{0}^{\star}(1430)$, and
$a_{0}(1450)~/~K_{0}^{\star}(800)$. Quite remarkably, the outcome of the fit
is univocal: only the pair $a_{0}(1450)~/~K_{0}^{\star}(1430)$ yields a good
fit, while the other combinations do not. We thus conclude that the $I=1$
and $I=1/2$ quark-antiquark scalar resonances lie above 1 GeV. 
In fact, the quality of our fit is surprisingly good.
As one can see in Fig.\ \ref{Fit} we describe all experimental quantities with 
an average error of 5\%, and most of them even to much better precision. 

We then study the two isoscalar mesons of the model. Setting
the (large-$N_c$ suppressed) parameters $\lambda_1$ and $h_1$ in
the Lagrangian \eqref{eq:Lagrangian} to zero,
these masses turn out 
to be about $1.36$ GeV and $1.53$ GeV, respectively. 
Our results for $f_{0}^{L}$ are in agreement with
the experimental decay widths of $f_{0}(1370).$ 
Our theoretical value for $f_{0}^{H}\rightarrow KK$ turns out to be too large, 
while $f_{0}^{H}\rightarrow\pi\pi$ vanishes, because $f_0^H\equiv \sigma_S$ is 
a pure $\bar{s}s$ state.
Nevertheless, our model predicts the existence of a scalar-isoscalar state which
decays predominantly into kaons; this is indeed the decay pattern shown by
$f_{0}(1710).$ For these reasons we identify our state $f_{0}^{H}$ as
(predominantly) $f_{0}(1710)$. It should be stressed that a quantitative 
study of the scalar-isoscalar system cannot be performed at present because 
our model does not contain the glueball state, the mass of which
is around 1.5 GeV and which
mixes with the two  $\bar q q$ scalar-isoscalar states, forming the three 
physical states in the 1.3--1.7 GeV energy region.

\section{CONCLUSIONS}

\label{sec:conclusion}

We have presented a linear sigma model with three flavors and global chiral
$U(3)_{L}\times U(3)_{R}$ symmetry. The model implements the symmetries of QCD,
the discrete $CPT$ symmetry, the global $U(N_{f})_{L}\times U(N_{f})_{R}$ 
chiral symmetry, and the breaking mechanisms of the latter symmetry.

The model includes meson states up to energies of $\sim 1.7$ GeV. 
This energy region exhibits numerous resonances, related by scattering 
reactions and decays. For this reason, a realistic model of QCD degrees of 
freedom in the mentioned energy region should describe as many of the 
resonances as possible. Thus, we have constructed a linear sigma model that 
contains scalar (two isoscalars,
$f_{0}^{L}$ and $f_{0}^{H}$, as well as an isotriplet, $\vec{a}_{0}$, and
two isodoublets, $K_{0}^{\star}$), pseudoscalar ($\pi$,
$K$, $\eta$, $\eta^{\prime}$), vector ($\rho$, $\omega$, $K^{\star}$, $\phi$),
and axial-vector [$a_{1}$, $K_{1}$, $f_{1}(1285)$, $f_{1}(1420)$] degrees of
freedom.

The model, dubbed extended Linear Sigma Model (eLSM), has allowed us
to study the overall phenomenology of mesons and, in particular, 
to explore the nature of
scalar and axial-vector resonances. In order to test our model we have
performed a global fit to 21 experimental quantities involving both the
(pseudo)scalar and the (axial-)vector masses and decays. Due to 
mixing with the scalar glueball (not included here), we did not
include the scalar-isoscalar resonances in the fit. 
Similarly, we have omitted the axial-vector resonance $K_{1},$ 
due to the fact that in reality a
large mixing of two kaonic fields from the $1^{++}$ and $1^{+-}$
nonets takes place.

One of the central questions of our discussions has been the assignment of the
scalar states: to this end we have tested the possible scenarios for the
isotriplet and isodoublet scalar states by assigning our scalar fields 
$\vec{a}_{0}$
to $a_{0}(980)$ or $a_{0}(1450)$ and $K_{0}^{\star}$ to $K_{0}^{\star}(800)$
or $K_{0}^{\star}(1430)$. The outcome is univocal: the global fit works well
only if the states $a_{0}(1450)$ and $K_{0}^{\star}(1430)$ are interpreted as
quark-antiquark states. On the contrary, the other combinations deliver large
and unacceptable values of $\chi^{2}$. We thus
conclude that the scalar $I=1$ and $I=1/2$ states have to be
identified with the resonances $a_{0}(1450)$ and $K_{0}^{\star}(1430).$
Moreover, the overall phenomenology described by the fit is very good, see
Ref.~\cite{eLSM}. 
The good agreement with data also shows that the axial-vector mesons can be
interpreted, just as their vector chiral partners, as quark-antiquark states.

We have then studied the consequences of our fit. We have primarily
concentrated on the scalar-isoscalar sector which was not included in the
fit. In the large $N_{c}$ limit it is possible to make clear
predictions for the two states $f_{0}^{L}$ and $f_{0}^{H}$. Their masses lie
above 1 GeV and their decay patterns have led us to identify $f_{0}^{L}$ with
(predominantly) $f_{0}(1370)$ and $f_{0}^{H}$ with (predominantly)
$f_{0}(1710).$ The theoretical decay rates of $f_{0}^{L}$ are in agreement
with experiment; $f_{0}^{H}$ decays only into kaons, but turns out to be
too wide. At a qualitative level, this large-$N_{c}$ result clearly shows that
also the scalar-isoscalar quark-antiquark states lie above 1 GeV. However, 
in this
system the inclusion of an additional glueball state would be necessary
for a full quantitative study. Mixing phenomena are known to be large here and
affect both the masses and decays. The study of a genuine 
three-state system is mandatory:
starting from $\sigma_N=\sqrt{1/2}(\bar{u}u+\bar{d}d),$
$G=gg$ and $\sigma_S =\bar{s}s$ to describe properly $f_{0}(1370),$
$f_{0}(1500),$ and $f_{0}(1710)$.

\begin{theacknowledgments}
Gy.\ Wolf and P.\ Kov{\'a}cs thank Goethe University for its hospitality.
They were partially supported by the Hungarian OTKA funds T71989 and T101438.
The work of D.\ Parganlija and F./ Giacosa was supported by the 
Foundation of the Polytechnical
Society Frankfurt. This work was financially supported by the 
Helmholtz International Center for FAIR within the framework of the 
LOEWE program 
(Landesoffensive zur Entwicklung Wisschenschaftlich-{\"O}konomischer 
Exzellenz) launched by the State of Hesse.
\end{theacknowledgments}

\bibliographystyle{aipproc}   


\end{document}